\documentclass{JAC2003}

\usepackage{graphicx}
\usepackage{booktabs}
\usepackage{amsmath}

\begin{document}
\title{BEAM COUPLING IMPEDANCE MEASUREMENT AND MITIGATION FOR A TOTEM ROMAN POT}

\author{M. Deile, F. Caspers, T. Kroyer, M. Oriunno, E. Radermacher,
A. Soter \\
(CERN, Geneva, Switzerland), F. Roncarolo (University of Manchester, UK).}

\maketitle

\begin{abstract}
The longitudinal and transverse beam coupling impedance of the first final 
TOTEM Roman Pot unit has been measured in the laboratory with the wire method. 
For the evaluation of transverse impedance the wire position has been kept 
constant, and the insertions of the RP were moved asymmetrically. 
With the original configuration of the RP, resonances with fairly high Q 
values were observed. In order to mitigate this problem, RF-absorbing 
ferrite plates were mounted in appropriate locations. As a result, all 
resonances were sufficiently damped to meet the stringent LHC beam coupling 
impedance requirements.
\end{abstract}

\section{THE TOTEM ROMAN POTS}
\begin{figure}[htb]
   \centering
  \includegraphics[width=0.49\columnwidth]{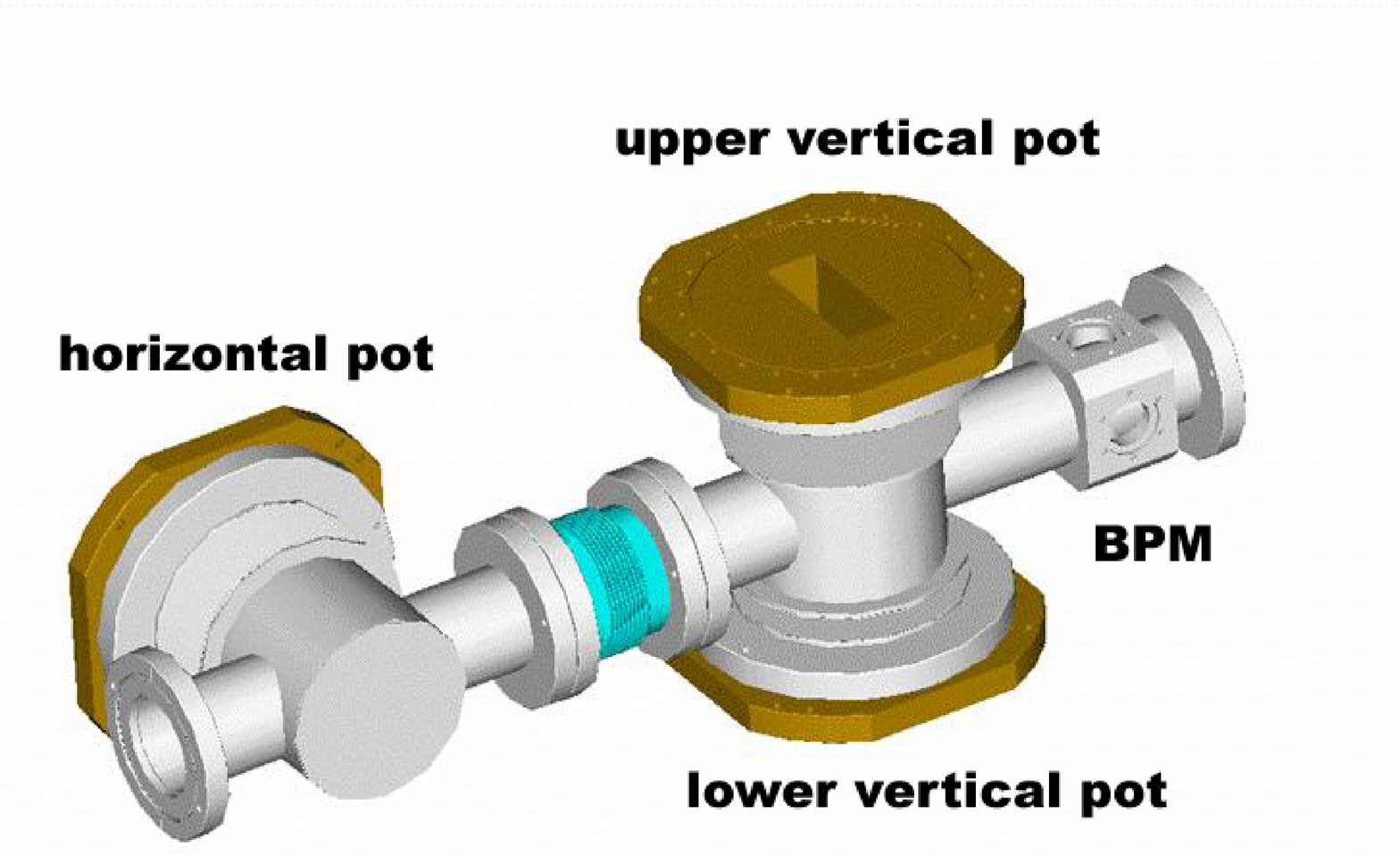}
  \includegraphics[width=0.49\columnwidth]{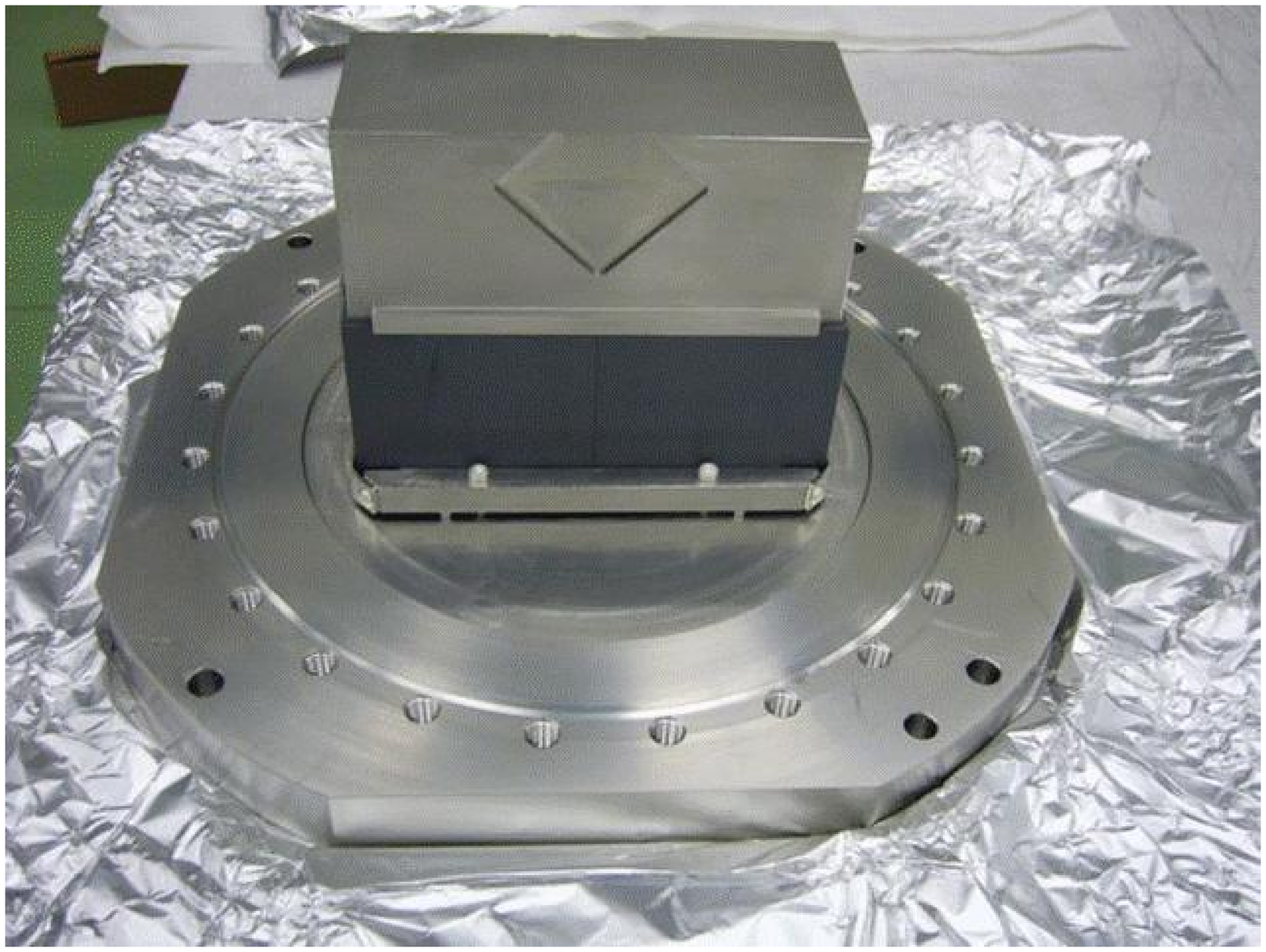}
  \caption{Left: the vacuum chambers of a RP unit
accomodating the horizontal and the vertical pots and a Beam Position 
Monitor. Right: the pot with the thin window and a Ferrite collar 
(black).}
  \label{fig:rp}
  \vspace*{-5mm}
\end{figure}

The LHC experiment TOTEM~\cite{tdr} is designed for measuring the 
elastic pp scattering
cross-section, the total pp cross-section and diffractive processes. These
physics objectives require the detection of leading protons with scattering
angles of a few $\mu$rad, which is accomplished using a Roman Pot (``RP'') 
system with stations at 147\,m and 220\,m from the interaction point 5 
where also CMS will be located.
Each station is composed of two RP units separated by a few metres
depending on beam equipment integration constraints.
Each RP unit consists of a vacuum chamber equipped with two 
vertical insertions (top and bottom) and a horizontal one (Fig.~\ref{fig:rp}).
Each insertion (``pot'') contains a package of 10 silicon detectors in a 
secondary vacuum. The pots can be moved into the 
primary vacuum of the machine through vacuum bellows. 
In order to minimise the distance of the detectors from the beam, 
and to minimise multiple scattering, the wall 
thickness of the pot is locally reduced to a thin window foil.

The low impedance budget of the LHC machine (broadband
longitudinal~impedance limit $Z/n \approx 0.1\,\Omega$) imposes a tight  
limit on the RPs' beam coupling impedance. 

\section{IMPEDANCE MEASUREMENT WITH THE WIRE METHOD}
\subsection{Longitudinal Impedance}

\begin{figure}[htb]
   \centering
  \vspace*{-2mm}
  \includegraphics[angle=-90,width=0.8\columnwidth]{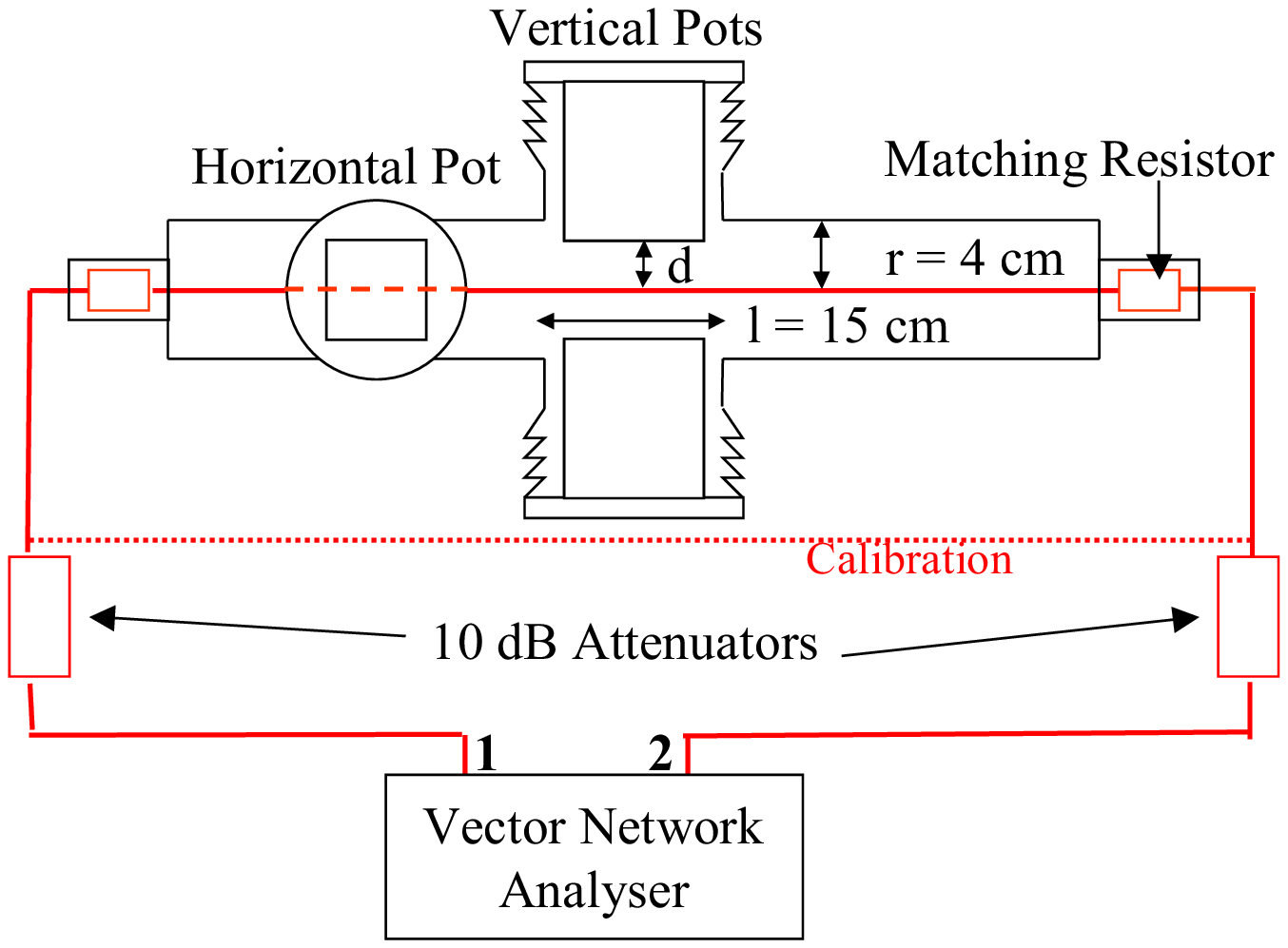}
  \caption{Setup of the impedance measurements.}
  \label{fig:setup}
  \vspace*{-2mm}
\end{figure}

The beam coupling impedance measurement was performed with the wire method
like with the first RP prototype in 2004~\cite{pac2005}.
After pulling a 0.3\,mm thick wire through the RP along its beam axis,
a vector network analyser was used to measure the complex
transmission coefficient $S_{21}(f,d_{x},d_{y})$ between the two ends of the RP
(Fig.~\ref{fig:setup}) as a function of the frequency $f$ 
and of the horizontal 
and vertical pot distances ($d_{x}$, $d_{y}$) from the wire.

\begin{figure}[htb]
   \centering
  \hspace*{-1mm}\includegraphics[width=0.91\columnwidth]{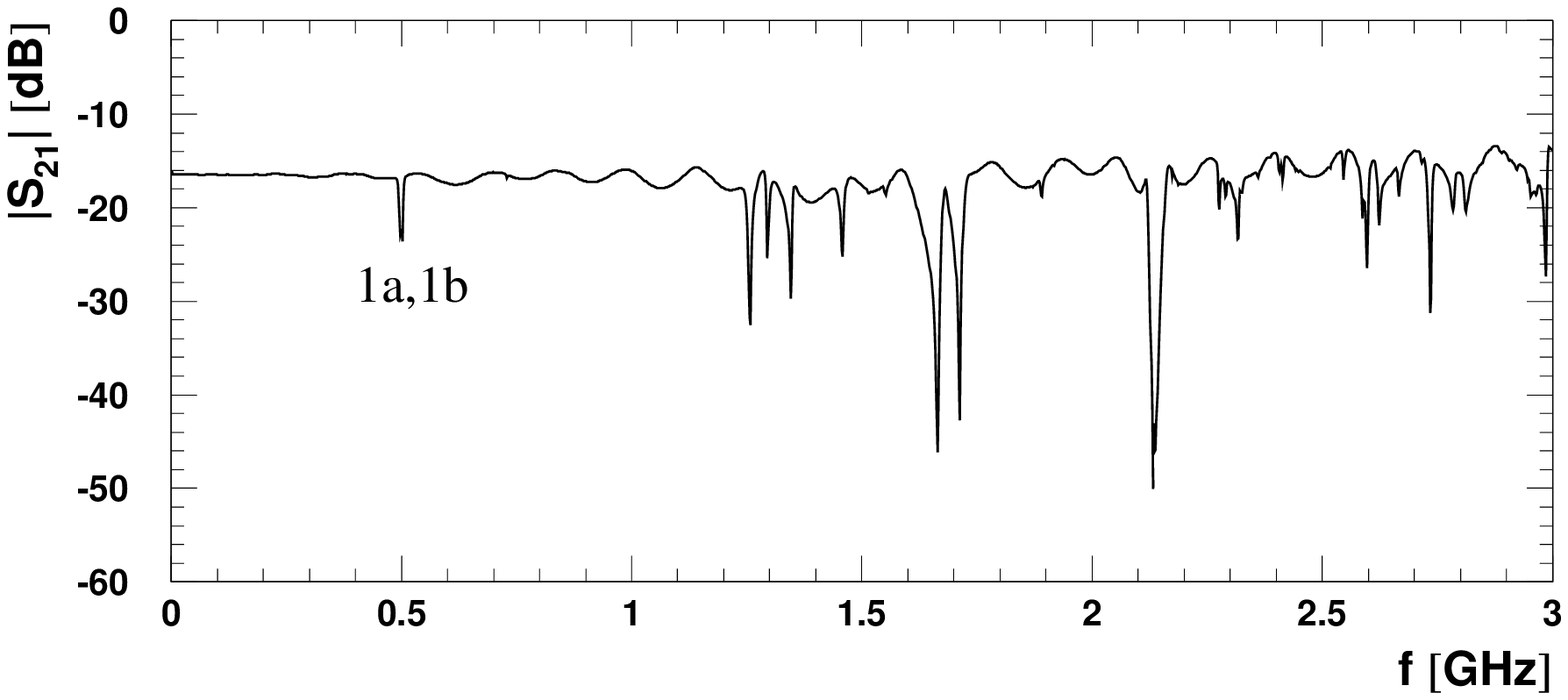}

  \vspace*{-12mm}
  \includegraphics[width=0.91\columnwidth]{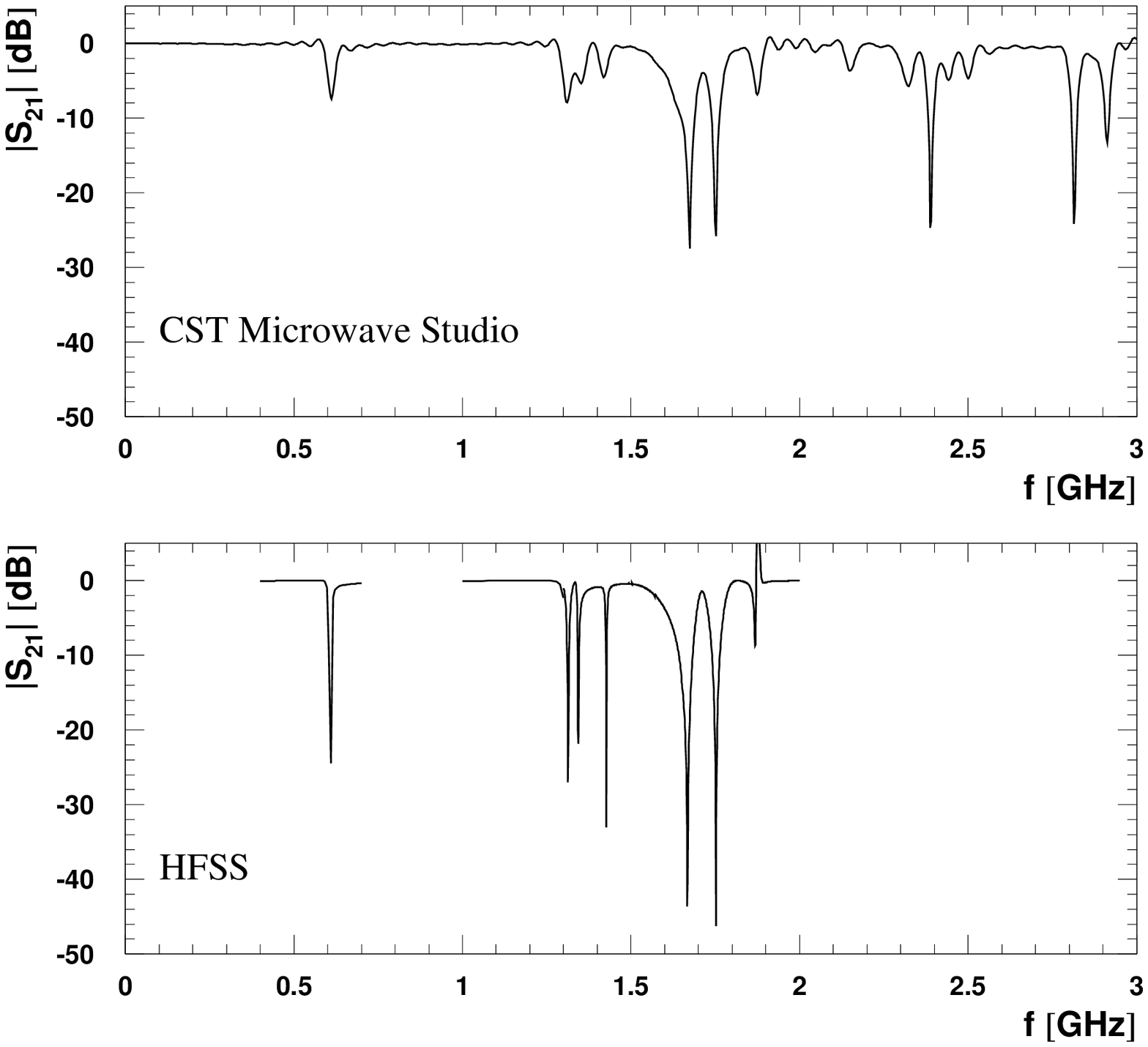}
  \vspace*{-7mm}
  \caption{Comparison of the measured (top) and the simulated 
(middle and bottom) transmission coefficient $|S_{21}|$ with all pots in 
retracted position ($d_{x}=d_{y}=40\,$mm).}
  \label{fig:spectra}
\end{figure}

Fig.~\ref{fig:spectra} shows the measurement result for all pots in retracted
position and compares it with simulations based on two different programs.
While the simulations describe qualitatively all main structures seen in the 
data, the resonances are shifted in frequency by up to 20\%. This 
disagreement is attributed to the modelling of the bellows in the simulations. 
An exact model requires a very dense mesh of the volume which compromises the 
simulation convergence in an acceptable CPU time. Substituting the bellows 
with a longer smooth cylinder of an equivalent total metallic surface 
provides good agreement at the first mode frequency ($\sim$500\,MHz), but 
does not succeed at higher frequencies. 
Improved numerical simulation models will be studied.

The longitudinal impedance $Z$ was calculated 
with the ``improved log formula''~\cite{jensen} 
\begin{multline}
Z(f,d_{x},d_{y}) = \\
= -2 Z_{C}\, \ln \frac{S_{21}(f,d_{x},d_{y})}{S^{ref}_{21}(f)} \left[ 1 + i\, \frac{\ln \frac{S_{21}(f,d_{x},d_{y})}{S^{ref}_{21}(f)}}{4 \pi l\,f / c} \right] ,
\end{multline}
where $Z_{C} = 294\,\Omega$ is the characteristic impedance of the
unperturbed beam pipe and 
$l = 15\,$cm is the length of the perturbation (i.e. the
diameter of the pot insertions).
The measurement with all pots in retracted position served as 
reference measurement $S^{ref}_{21}(f)$ after removal of all resonances by 
interpolation in both modulus and phase.

\begin{figure}[tb]
   \centering
  \includegraphics[width=\columnwidth]{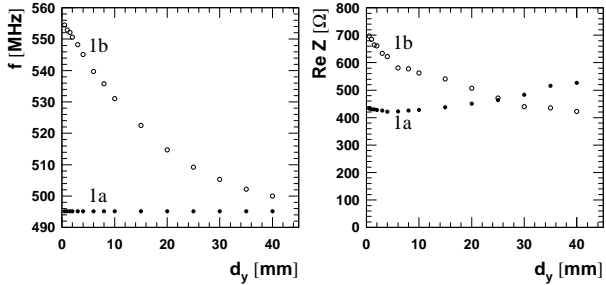}
  \vspace*{-5mm}
  \caption{Frequency and impedance of the first double resonance as a function
of the distance of the vertical pots from the wire, with retracted horizontal
pot.}
  \label{fig:resonance_vs_d}
\end{figure}
\begin{figure}[htb]
   \centering
  \includegraphics[width=\columnwidth]{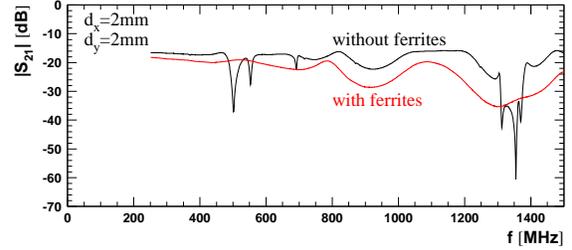}
  \caption{Frequency spectrum of the transmission coefficient $|S_{21}|$ 
before and after installation of the ferrite tiles.}
  \label{fig:ferrites}
\end{figure}

The approximately Gaussian LHC bunch 
structure with $\sigma_t = 0.25\,$ns leads to a Gaussian envelope with 
$\sigma_f = 0.63\,$GHz in the 
frequency distribution of the LHC current with harmonics every 40\,MHz.
Hence the relevant resonances lie well below 1\,GHz. 
Fig.~\ref{fig:resonance_vs_d} shows the evolution of frequency and 
longitudinal impedance 
of the first double resonance peak as a function of the vertical pot position
$d_{y}$ (for this measurement the top and the bottom pot positions were 
symmetrical w.r.t. the wire). The impedance value of 700\,$\Omega$ for 
mode 1b at $d_{y} = 0.5\,$mm corresponds to
$Z/n = 14\,{\rm m}\Omega$, where 
$n = f_{reson}/f_{LHC} = 555\,{\rm MHz} / 11\,{\rm kHz}$,
or a dissipated power of about 200\,W, which demonstrates the necessity of 
mitigating hardware modifications. The latter were realised in the form of 
a collar of ferrite tiles around the pot insertions (Fig.~\ref{fig:rp}, right).
They had the desired effect of smearing out all resonances beyond recognition
(Fig.~\ref{fig:ferrites}).

\subsection{Transverse Impedance}
\begin{figure}[htb]
   \centering
  \includegraphics[width=\columnwidth]{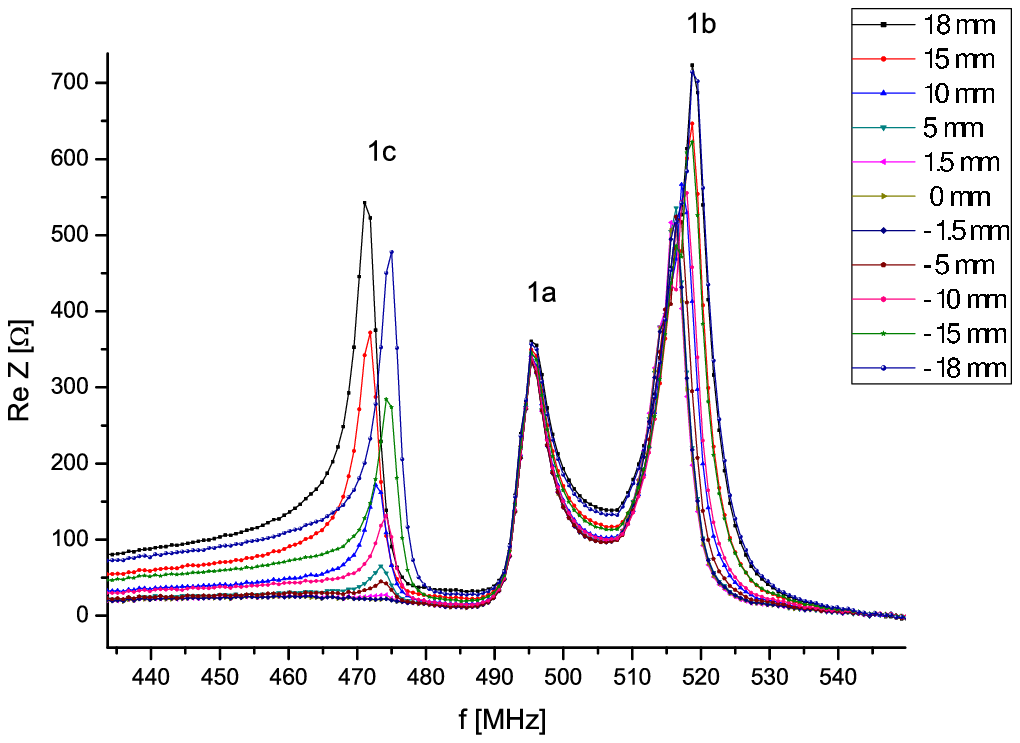}
  \includegraphics[width=\columnwidth]{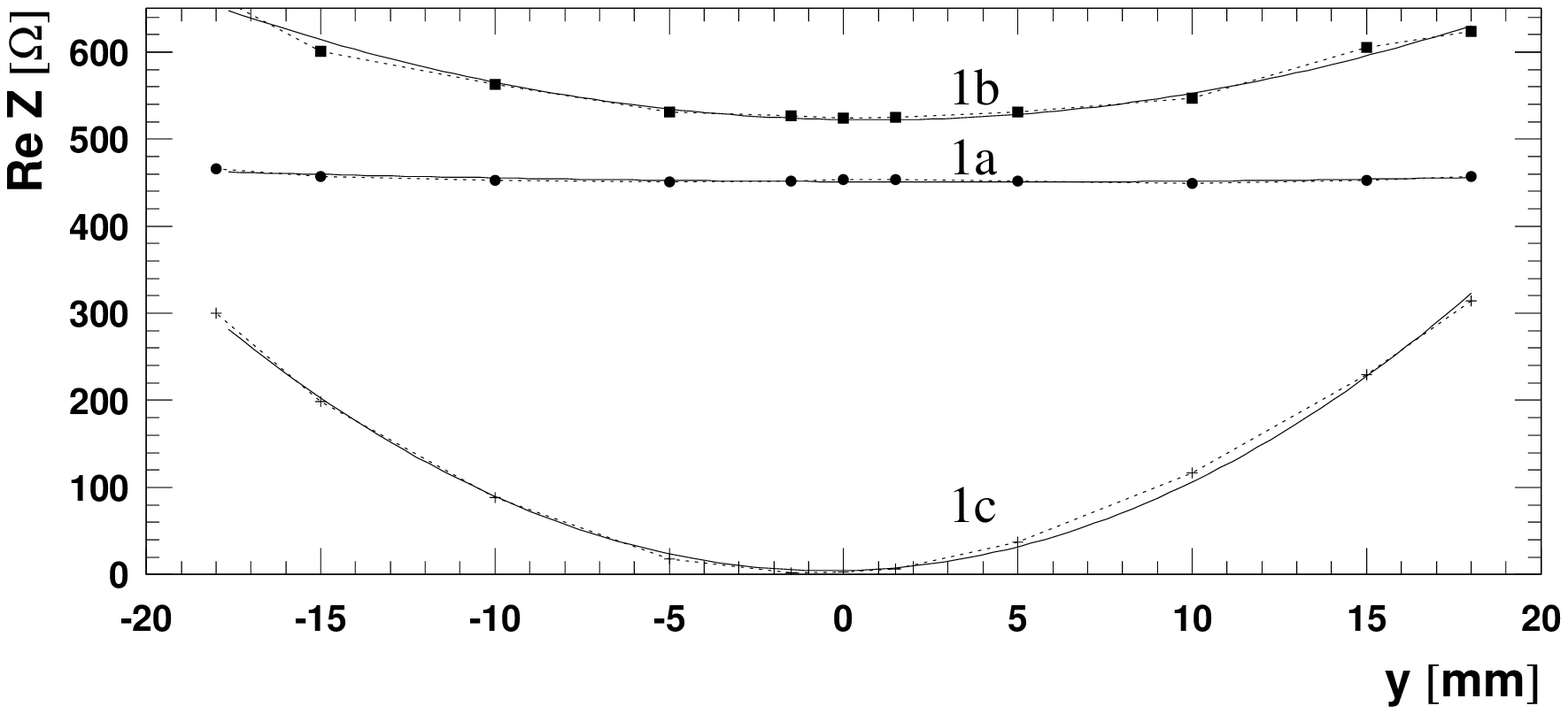}
  \includegraphics[width=\columnwidth]{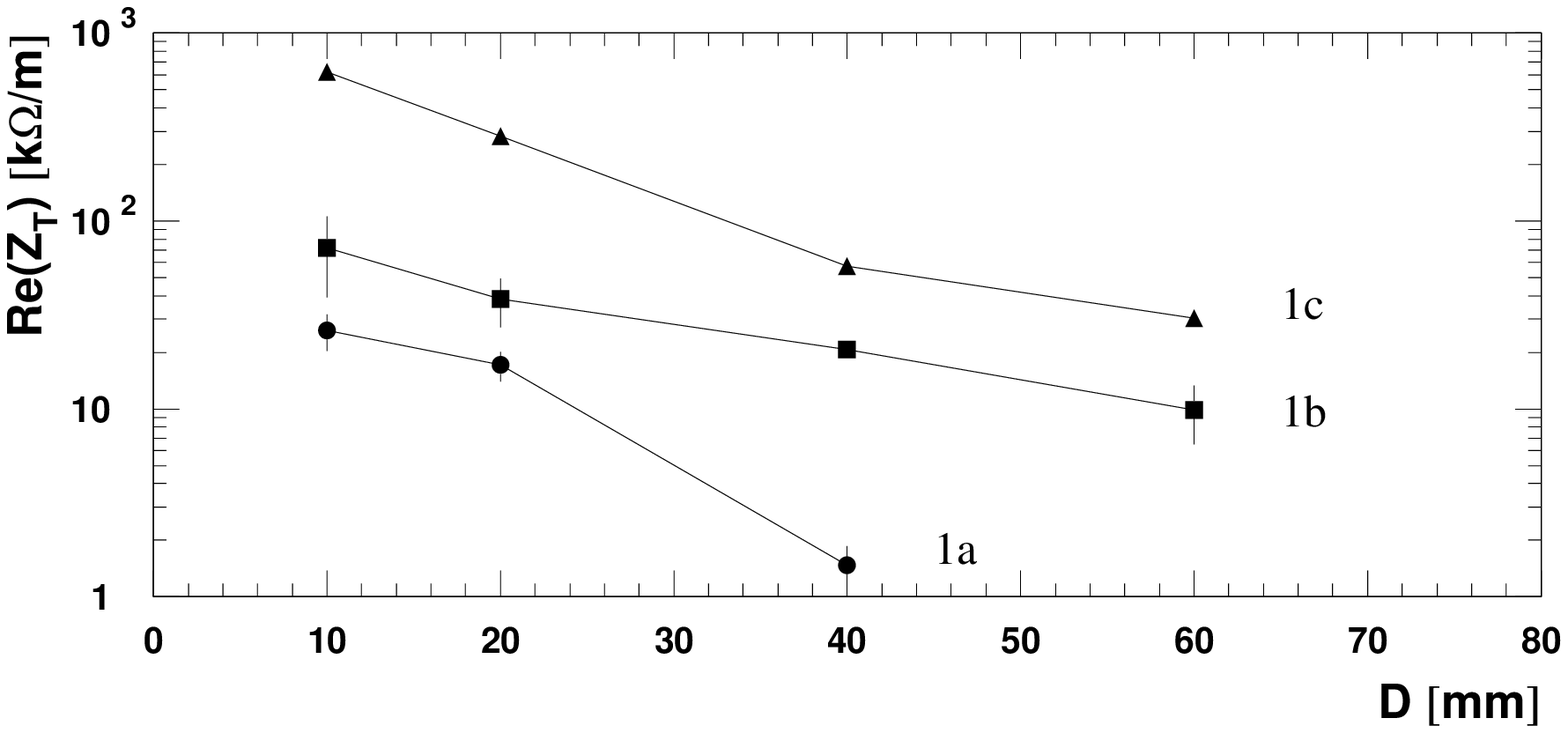}
  \caption{Study of the transverse impedance of the first resonance group.
Top: Frequency spectra around the first resonance group for a fixed vertical 
RP aperture width $D =\,$40\,mm and different excentricities $y$ 
of this aperture with respect to the wire;
middle: longitudinal impedance as a function of the excentricity $y$
for the fixed aperture width $D =\,$40\,mm; 
bottom: dependence of the transverse impedance on the aperture width $D$.
For this entire study, the horizontal pot was in retracted position.}
  \label{fig:transverse}
\end{figure}

The transverse impedance was only measured for the configuration without
ferrite tiles where resonances were visible. For technical reasons, neither
the two-wire method nor a movable wire were practicable. Instead, the two
vertical pots were moved asymmetrically, keeping their relative distance $D$ --
the jaw width -- constant. Then the longitudinal impedance was measured as
a function of the excentricity $y$, i.e. the position of 
the jaw centre with respect to the wire (Fig.~\ref{fig:transverse}, top and
middle). After a parabolic fit 
$Z_{L}(y) = z_{0} + z_{1} y + z_{2} y^{2}$, where ideally 
$z_{1} = 0$ by virtue of symmetry, a
combination of vertical transverse impedance $Z_{Ty}$ and detuning impedance
$Z_{det}$ can be obtained from the curvature parameter $z_{2}$,
like in the moving wire technique~\cite{abnote}:
\begin{equation}
Z_{Ty} + Z_{det} = \frac{c}{2\,\pi\,f} z_{2} \:.
\end{equation}
Since there is only one horizontal pot, no analogous measurement of the 
x-component $Z_{Tx} - Z_{det}$ could be made, which would have enabled the 
elimination of $Z_{det}$. However, based on the calculations in 
Ref.~\cite{tsutsui} for different aperture geometries, the contributions 
$Z_{Ty}$ and $Z_{det}$ could be approximately disentangled.
The result for the first group of resonances is shown in 
Fig.~\ref{fig:transverse} (bottom) as a function of the jaw width $D$.

\subsection{Time Domain Studies -- the Loss Factor}
\begin{figure}[htb]
   \centering
  \includegraphics[width=\columnwidth]{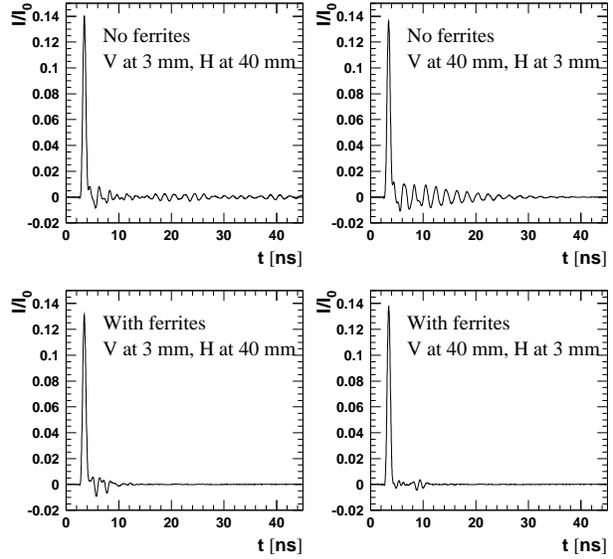}
  \caption{Transmission of a Gaussian pulse of magnitude $I_{0}$ 
without (top) and with (bottom)
ferrites for two different combinations of the horizonal (H) and vertical (V)
pot distances from the wire.}
  \label{fig:time}
\end{figure}

The built-in Fourier transformation capability of the network analyser 
facilitated a transmission study in the time domain. Fig.~\ref{fig:time} 
shows the transmission response to an injected Gaussian pulse with 
$\sigma = \,$0.6\,ns for two different position configurations of the 
horizontal and vertical pots. While without ferrites the resonant behaviour
of the insertion cavities 
leads to oscillations extending beyond 25\,ns after the main pulse, this
``ringing'' is suppressed by the ferrites after less than 10\,ns.

\begin{figure}[htb]
   \centering
  \includegraphics[width=\columnwidth]{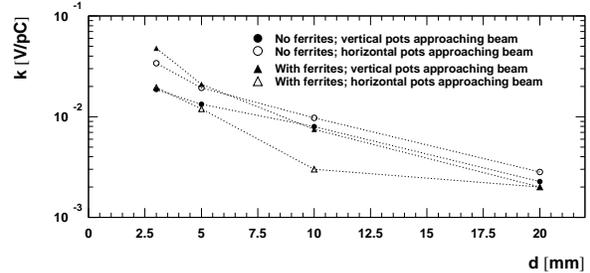}
  \caption{Loss factor as a function of the pot distance from the wire
with and without ferrites, distinguishing movements of the vertical and 
horizontal pots.}
  \label{fig:lossfactor}
\end{figure}

Focussing on the main pulse, the loss factor can be 
calculated according to the formula~\cite{sands}
\begin{equation}
k(d_{x},d_{y}) = 2\,Z_{C}\,
\frac{\int{I_{ref}(t) \left[I_{ref}(t) - I(d_{x},d_{y},t)\right] dt}}
{\left[\int{I_{ref}(t)\, dt}\right]^2} \:,
\end{equation}
where the time integral extends over a range of $\pm 2\sigma$ around the peak.
$I_{ref}(t)$ is the reference pulse measured with all pots in 
retracted position. As Fig.~\ref{fig:lossfactor} shows, the loss factor does
not change strongly with the addition of the ferrites.

\section{Acknowledgements}
We thank J. Noel, A.G. Martins de Oliveira and S. Rangod 
for the assembly of the Roman Pot.

\end{document}